\documentclass[12pt,preprint]{aastex}
\usepackage{amsmath}

\shorttitle{Variability in the WISE PDR}
\shortauthors{Hoffman, D.I., et al.}

\begin{document}

\title{Variability Flagging in the WISE Preliminary Data Release}

\author{D.I. Hoffman\altaffilmark{1}, R.M. Cutri\altaffilmark{1}, F.J. Masci\altaffilmark{1}, J.W. Fowler\altaffilmark{1}, K.A. Marsh\altaffilmark{1}, and T.H. Jarrett\altaffilmark{1}}
\altaffiltext{1}{Infrared Processing and Analysis Center,\\ California Institute of Technology,\\ Pasadena, CA 91125, USA; dhoffman@ipac.caltech.edu}

\begin{abstract}
The Wide-field Infrared Survey Explorer Preliminary Data Release Source Catalog contains over 257 million objects.  We describe the method used to flag variable source candidates in the Catalog.   Using a method based on the chi-square of single-exposure flux measurements, we generated a variability flag for each object, and have identified almost 460,000 candidates sources that exhibit significant flux variability with greater than $\sim 7 \sigma$ confidence.  We discuss the flagging method in detail and describe its benefits and limitations.  We also present results from the flagging method, including example light curves of several types of variable sources including Algol-type eclipsing binaries, RR Lyr, W UMa, and a blazar candidate.

\end{abstract}

\section{INTRODUCTION}   
The Wide-field Infrared Survey Explorer \citep[WISE; ][]{Wright10},
mapped the entire sky in four bands centered at wavelengths of 3.4,
4.6, 12, and 22 $\mu$m (hereafter W1, W2, W3, and W4).  WISE conducted
its survey using a 40 cm cryogenically cooled telescope equipped with
four 1024x1024 infrared array detectors that simultaneously imaged the
same 47'x47' field-of-view on the sky.  WISE flew in a 531 km
sun-synchronous polar orbit and employed a freeze-frame scanning
technique whereby the telescope scans the sky continually at a rate
of approximately 3.8 arcmin/sec and a scan mirror freezes the sky on
the focal plane detectors while 7.7 sec (W1 and W2) and 8.8 sec (W3
and W4) exposures are acquired.  The FOV of each successive exposure
set overlaps the previous one by 10\%, and the scan paths on adjacent
orbits overlap by approximately 90\% on the ecliptic.  The WISE survey
strategy alternated stepping the scan path forward and backward on
subsequent orbits in an asymmetric pattern that approximately matched
the orbital precession rate.  In this way, each point near the
ecliptic was observed on every other orbit, approximately each 191
minutes, and typically 12 independent exposures were accumulated for
each point near the ecliptic.  The number of exposures increases with
ecliptic latitude, reaching over 1000 at the ecliptic poles.

The WISE observing cadence is well-suited for studying periodic variable stars
with periods less than $\sim 2$ days near the ecliptic.  The most common
variables in this category are Algol, $\beta$ Lyrae, and W Ursae
Majoris (W UMa) -type eclipsing binaries, and RR Lyrae pulsating variable
stars.  W UMa variables are the most numerous among these, as ~95\% of all
variable stars in the solar neighborhood are W UMa stars
\citep{eggen67}, but their mid-infrared emission is relatively weak,
and their variation amplitudes can be small and difficult to detect.
Therefore, Algol and RR Lyr variables are the most common periodic variables
detected by WISE.

Non-periodic variables are also detected by WISE.  Among these are
young stellar objects (YSOs) such as those seen in Orion and other
star formation complexes, and active galactic nuclei (AGN).  Blazars
are be the most common type of AGN with variability detected by WISE
because of the short timescale variations they exhibit \citep{stein76}.
Other types of AGN vary in brightness over longer timescales, and
therefore it is difficult to probe their variability with WISE,
except perhaps near the ecliptic poles where the coverage period is
longest.  As we show below, AGN can be separated from the other types
of variables using WISE colors in the cases where the source has a 4.6
$\mu$m detection, and with more confidence when a 12 $\mu$m detection is available.

The WISE Preliminary Data Release Source Catalog \citep{cutri11}
contains entries for over 257 million objects that were detected on
the coadded WISE Atlas Images covering ~57\% of the sky.  Photometry
for the Source Catalog entries was performed by fitting point spread
functions (PSFs) simultaneously to all of the individual exposures
covering an object.  The magnitudes reported for each object
effectively average over any time variations, although measurements of
the flux variance across all individual exposures are available from
which we can test for variability.  Each source record is accompanied
by a variability flag, $var\_flg$, that is related to the probability
of flux variation based on the distribution of source flux
measurements.

In this paper, we describe the derivation of the WISE Source Catalog
variability flagging, and demonstrate how to apply the variability
flag to identify true variable sources.  Examples of light curves for
various classes of objects are presented, as well as some
representative phased light curves for periodic variables with large
numbers of measurements.

\section{Variability Flagging Method}

The method used to identify the variables makes use of statistics computed in the multi-frame data processing.  Traditional variability indices such as the Stetson $L$, $J$, and $K$ indices \citep{welch93,stetson87,stetson96} were not used.  The main reason is that access to the single-exposure flux measurements is not practical on such a large scale, whereas statistics from the multi-frame generation step are readily available.  The Stetson indices are also rather insensitive to transient events, such as flare stars, although the method we used also suffers from this problem.  WISE was not designed to identify variables, so using the multi-frame statistics gives the most practical and useful variability information. 

During the WISE multi-frame processing, fluxes are measured from all individual exposures simultaneously at the location of the source detection in the coadded image.  The standard deviation of the ensemble fluxes are computed and included in the WISE source catalog.  From these data, a variability flag is generated ($var\_flg$), which is an integer value (0-9) for each WISE band.  A value of zero indicates insufficient or inadequate data to determine a variability probability, while values of 1-9 indicate increasing likelihood of variability. 

As a first step, measurements are pre-filtered to eliminate sources that are not suited to make a variability determination.  The following lists the pre-filter conditions used and the definitions and/or reasoning behind each.  Each pre-filter was based on constraints on columns from the WISE Preliminary Release Source Catalog, with the exception of the single-frame SNR proxy.  See Table~\ref{tab:defs} for a brief description of the columns discussed in this paper.  More detailed definitions can be found in \citet{cutri11}.  In the conditions below and in Table~\ref{tab:defs}, ``?'' denotes a WISE band, and is equal to 1,2,3 or 4.

\begin{itemize}
\item {\boldmath $\frac{w?nm}{w?m} > 0.8$} : Reliability of single-frame measurements. This helps to ensure there is a detection of the source in the single-exposure image. These parameters ensure at least 80\% of the frames have a single-exposure detection at the location of the source in the multiframe image. However, this does not ensure there are also single-exposure flux detections for the source, as the single-exposure and multi-frame fluxes are measured at slightly different positions.

\item {\boldmath$w?rchi2 < 5.0$} : This reduces confusion issues by rejecting sources with high template-fit chi-square values. High chi-square values normally indicate confusion with nearby objects, or that the source is extended, both of which are conditions that can lead to measurement contamination.

\item {\bf {\boldmath $cc\_flags = 0$} in a least one band} : Contamination by image artifacts, especially diffraction spike artifacts, can produce spurious variability since the artifact location and intensity can vary between frames. Variability flagging is unreliable in the presence of artifacts and is avoided.

\item {\boldmath$w?m > 10$} : Small depth of coverage produces unreliable statistics since outliers have a larger effect on the overall variability. This filter helps to ensure that any variability is robust against noisy measurement outliers, by requiring a minimum depth of coverage.

\item {\bf Single-frame SNR proxy {\boldmath $> 5$}} : This helps to ensure there is a single-frame detection.  See the more detailed explanation later in this section.

\item {\boldmath$w?snr > 5$} : The coadd photometry SNR must be significant, otherwise there will be no single-exposure detections. This filter ignores sources that are too faint to have any single-exposure detections.

\item {\boldmath$na=0$} : Eliminates sources with active deblending, which indicates one more more close neighbors that can produce false variability.

\item {\boldmath$nb < 3$} : Limit to the number of blend components to two.  This also avoids confusion, which can generate variability. Tests showed that some known variables were still detected with $nb = 2$, so the threshold was set at this value. For sources with $nb > 2$, spurious flux variations may be seen due to confusion.

\item {\bf {\boldmath$w?sigmpro$} not null in at least one band} : There must be a multi-frame detection of the source.

\item {\bf {\boldmath$w?sigp1$} not null in at least one band} : There must be single-frame measurements for the source.

\end{itemize}

The single-frame SNR proxy is needed to ensure the variability flag is applied to actual single-exposure source detections.  The single-frame SNR cannot be estimated simply by using the value $w?snr / \sqrt{w?m}$, as the multi-frame SNR estimation employs a noise model based on zero-mean random fluctuations and a systematic component due to PSF error.  The latter does not average out over multiple measurements, so scaling the noise variance by the inverse of the number of measurements is not applicable. To adjust for this, approximately 100,000 random sources were used to compile the actual SNR of both single- and multi-frame measurements.  A linear fit was then applied to the function
\begin{equation}
 Y(w?mpro) = \sqrt{w?m} \frac{w?snr_{se}}{w?snr_c}
\end{equation}
where $w?mpro$ is the coadd magnitude of the source for a given band, $w?m$ is the number of frames that went into the multi-frame image, $w?snr_{se}$ is the SNR of the single-exposure measurement, and $w?snr_c$ is the SNR of the coadded source.  The result is that 
\begin{equation}
 SNR_{se} \approx \frac{Y(w?mpro) \cdot w?snr_c}{\sqrt{w?m}}
\end{equation}
gives a reasonable estimation of the single-frame SNR as a function of magnitude, given only the SNR of the coadd.

An estimate of the population standard deviation, $\sigma$,
is determined in each band by taking the $65^{th}$ percentile of
$w?sigp1$ in magnitude bins of 0.50.  The $65^{\rm{th}}$ percentile
was used instead of the median for conservatism and to ensure that
candidate variable sources in the highest confidence level bins do not
have many spurious variable sources mostly arising from unflagged
artifacts. These distributions form a ``null'' non-variable
distribution from which approximate which chi-squares can be calculated.  These
distributions are saved as a look-up table as functions of M and
magnitude and later interpolated upon retrieval. There is an effective
faint-magnitude cutoff at 16.0, 14.75, 10.75, and 7.0 for bands 1,2,3,
and 4, respectively. This is due to the lack of usable data at these
fainter magnitudes to make the look-up table, as these magnitudes are
approximately the single-exposure detection limit. Sources fainter
than these magnitudes will always receive a variability flag value of
"0" in the appropriate band.  At the bright end, there was also a
general lack of usable data as sources start to saturate and have
greater measurement errors as well as large chi-square values.  The
bright end of the look-up table ends at 6.5, 6.0, 2.5, and 1.5
magnitudes for bands 1, 2, 3, and 4, respectively.  Linear
interpolation was used for sources brighter than these limits, which
imposes a higher than normal $\sigma$ for the calculation with these
sources.  This ensures that very bright sources are not flagged as
variable in error due to sparse data and high measurement errors due
to saturation.  However, very bright sources flagged as variable
should still be treated with suspicion.

To generate the flag, we construct a function that is monotonically related to the likelihood that the null hypothesis is false, where the null hypothesis is that the observed fluxes came from a single real non-variable source. If the noise in the data were uncorrelated and Gaussian-distributed, the ideal function would be chi-square computed from the sample of fluxes.  The chi-square statistic is given by
\begin{equation}\label{eq:chi2}
\chi^2 = \sum_{i=1}^M\left(\frac{X_i - \mu}{\sigma}\right)^2
\end{equation}
where $M$ is the number of measurements (equivalent to $w?m$), $X_i$ is the single-exposure flux measurement for frame $i$, $\mu$ is the flux population mean,  and $\sigma^2$ is the flux variance of the non-variable population.  This equation cannot be used as is, because the individual flux measurements are no longer available.  The sample variance of the flux measurements, however, is available and given by $w?sigp1$:

\begin{equation}
w?sigp1^2 = \frac{1}{M}\sum_{i=1}^M\left(X_i - \mu\right)^2
\end{equation}
Substituting $w?sigp1^2$ in Equation~\ref{eq:chi2} and multiplying by $w?m$, the equation reduces to

\begin{equation}\label{eq:chifinal}
\chi^2 = \frac{w?m \cdot (w?sigp1)^2}{\sigma^2}
\end{equation}
The number of degrees of freedom, $N$, is given by $N = (w?m -1)$.

There are three significant approximations in Equation \ref{eq:chifinal} that cause
some deviation from the behavior of a true chi-square
statistic. Appendix A expands on some of these approximations.  

First, it operates in the magnitude domain, whereas it is
the flux domain that is best modeled as Gaussian. For sources with
high SNR, the magnitude distribution is also well approximated as
Gaussian, but for low SNR, the magnitude density function has an
appreciable tail, positive skewness, and positive excess kurtosis
(e.g., at SNR = 3, the skewness is 1.55, and the excess kurtosis is
3.63; at SNR = 10, these are down to 0.31 and 0.21,
respectively). This causes the chi-square approximation to have
excessive variance for low SNR sources. 

Second, in some cases the sample variance has been subjected to outlier trimming during the production processing. This appears to affect primarily bright sources and results in reconstructing the chi-square estimate with an overestimated number of degrees of freedom.

Third, although correlated errors do not affect the expectation value
of the sample variances, ignoring them results in a significant
inflation of the variance of the reconstructed chi-square defined in
Equation 5, making it considerably less efficient than a true
chi-square but still statistically a monotonic function of the
likelihood of the null hypothesis and therefore useful for our
objective. Using it with the chi-square $Q$ distribution propagates this
approximation error, but as discussed below, the value of $Q$ is
presented with very low precision, namely a one-digit value for
$-\rm{log}_{10} \rm{ }Q$, and so errors of a factor of 3 in $Q$ operate at the
truncation-error level. We require extreme statistical significance
for variability tagging, so even if $-\rm{log}_{10} \rm{ }Q$ is off by as much as $-3$, it is
still useful as an indicator that the null hypothesis should be
rejected. To calibrate an empirical $Q$ function for the actual
magnitude data is beyond the present scope, and so we use the
chi-square $Q$ function for this purpose.

\begin{figure} [hbtp]
  \centering \includegraphics[width=0.7\textwidth]{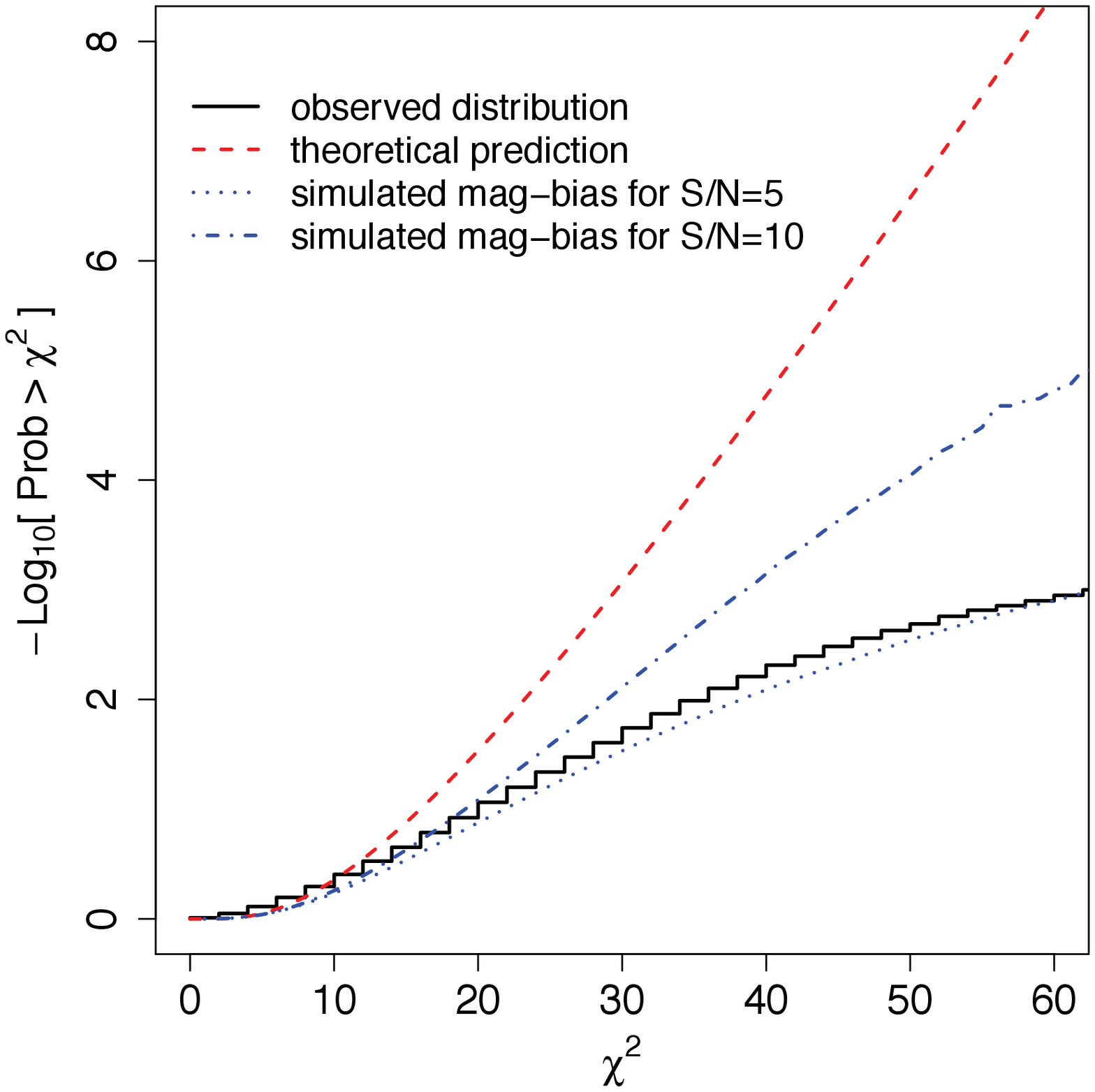}
    \caption[]  
{Cumulative distribution of the empirical $-\log_{10}Q$ values (stepped histogram) for a depth-of-coverage of $M = 11$ and the W1 magnitude range: $11 \leq m \leq 15$, where $Q$ = probability $>\chi^2$. Also shown is the expected trend of $-\log_{10}Q$ predicted from a $\chi^2$ distribution for $M-1 = 10$ degrees-of-freedom (dashed). The dotted and dash-dotted curves are from Monte-Carlo simulations to illustrate the impact of using magnitude units (instead of fluxes) on the noise-variance for two signal-to-noise ratios. The observed distribution was generated by a subset of the preliminary data release, consisting of approximately 6.6 million sources.  See text for details.}
  \label{fig:logQvsChi}
\end{figure}

Because of the preponderance of faint sources, the dominant
approximation error appears to stem from the non-Gaussian magnitude
distribution. Figure \ref{fig:logQvsChi} shows the observed distribution of $-\rm{log}_{10} \rm{ }Q$
as a function of chi-square for the case of 10 degrees of freedom and
W1 magnitudes of $13\pm2$. The observed distribution is very close to the
simulated case for SNR = 5, which included the effects of the
magnitude distribution being non-Gaussian but did not include the
effects of ignoring correlations in the photometric errors.

The probability, $Q$, that $\chi^2$ would be at least as large as what was observed given the null hypothesis, is calculated from the integral of the probability
density function:
\begin{equation}
Q = 1-\left[ \int_0^{\chi^2} \frac{1}{2^{N/2}\Gamma(N/2)} x^{(N/2 -1)} {\rm e}^{-x/2}dx \right]
\end{equation}
where $\Gamma$ is the gamma function.  The value of the band component of the variability flag, $F$, is then defined as the floor of the negative logarithm of $Q$:
\begin{equation}
F =  \rm{floor}\left[ -\rm{log}_{10}\rm{ } Q \right]
\end{equation}
with $F$ clipped at 9 for $F > 9$ and set to 1 when $F = 0$.  The final variability flag, $var\_flg$, is a four character string comprised of the values of $F$ for each of the four WISE bands.

\begin{figure} [hbtp]
  \centering \includegraphics[width=0.7\textwidth]{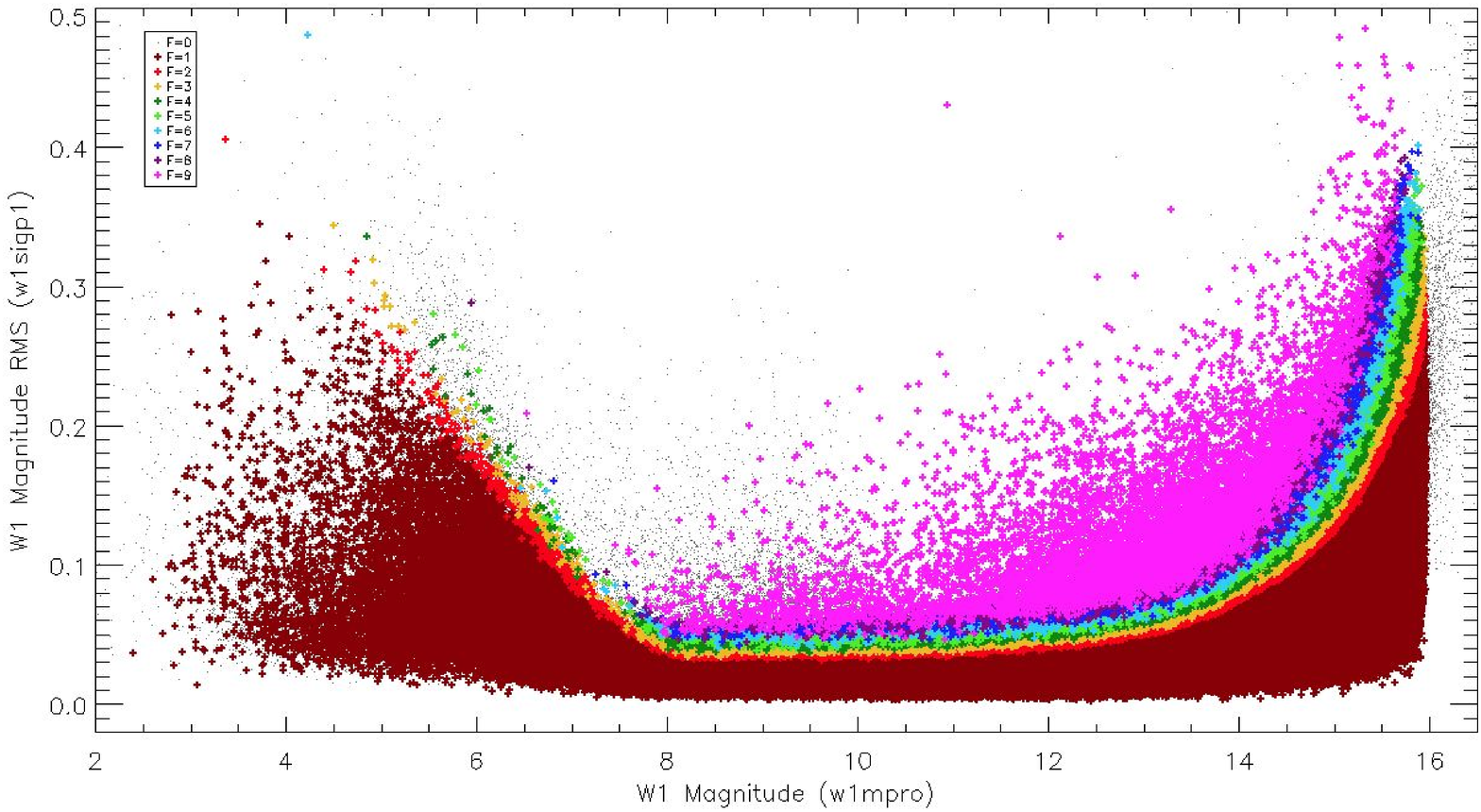}\includegraphics[width=0.3\textwidth]{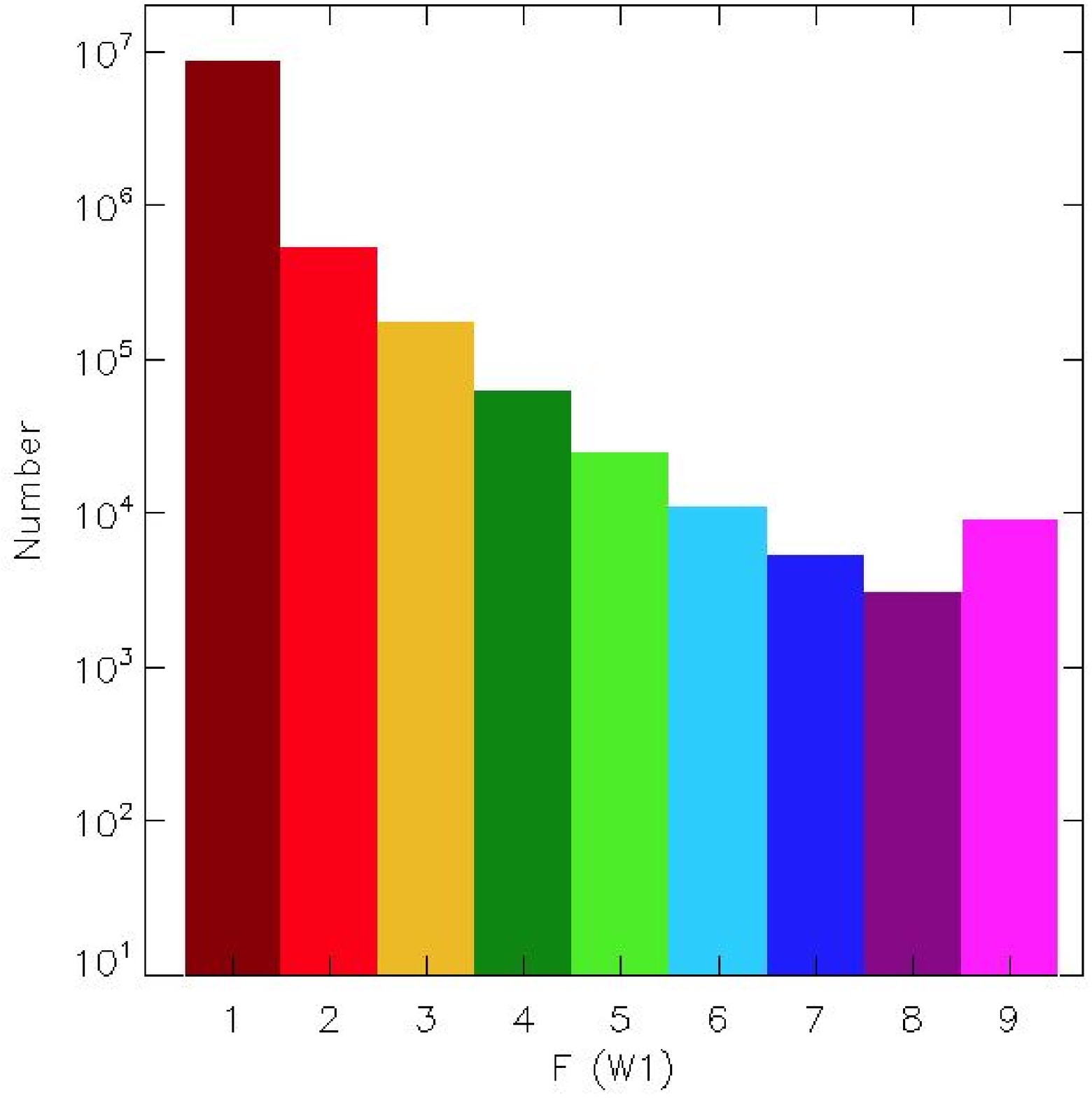}
    \caption[]  
{The standard deviation of the single-frame measurements as a function of the W1 multi-frame magnitude (left panel).  Each colored cross corresponds to a different $F$ value.  Grey dots are sources with $F=0$, which lack sufficient data for a classification. The right panel is a histogram of the different $F$ values with the appropriate bar color for all data in the left panel.}
  \label{fig:w1f}
\end{figure}

\begin{figure} [hbtp]
 \centering \includegraphics[width=0.7\textwidth]{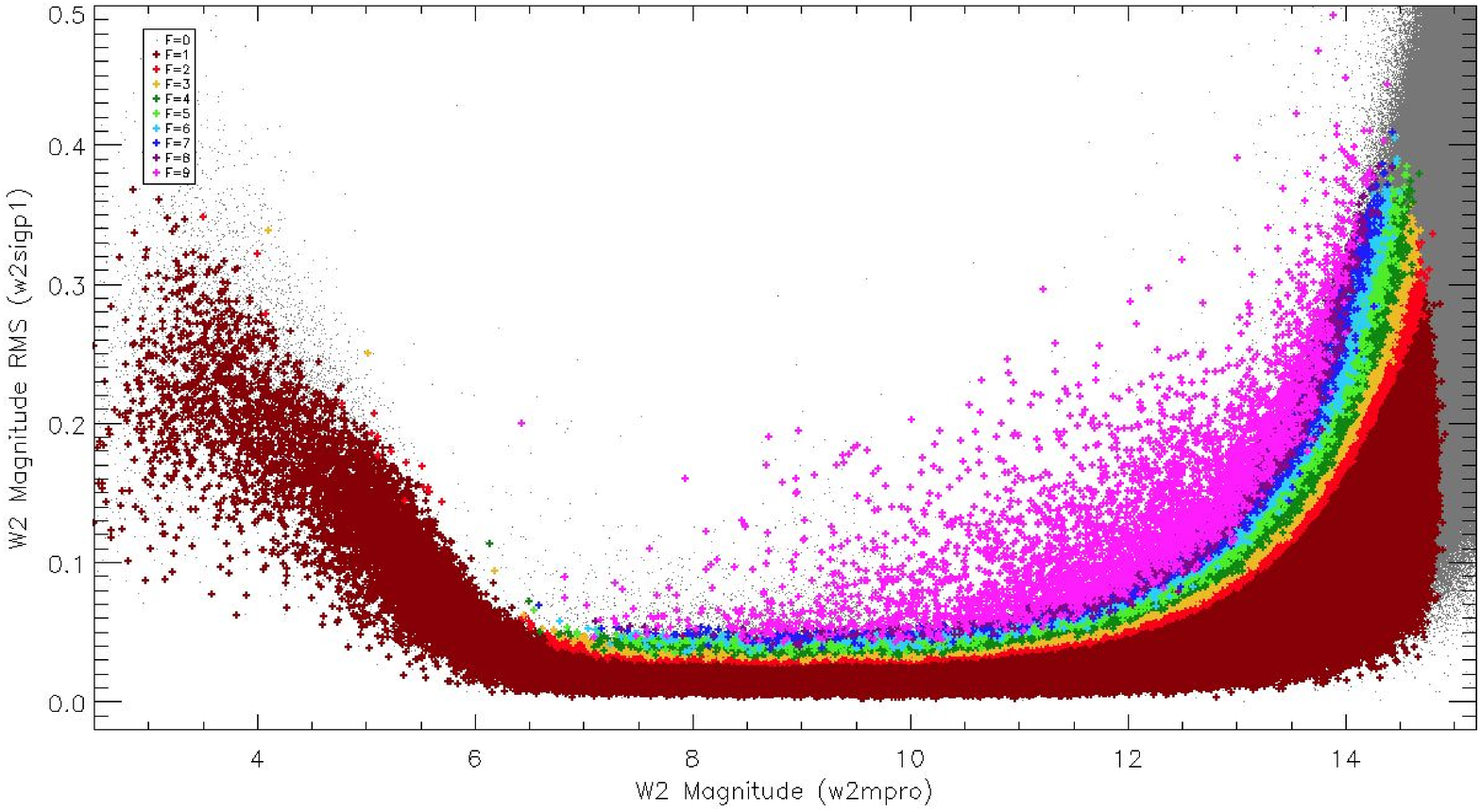}\includegraphics[width=0.3\textwidth]{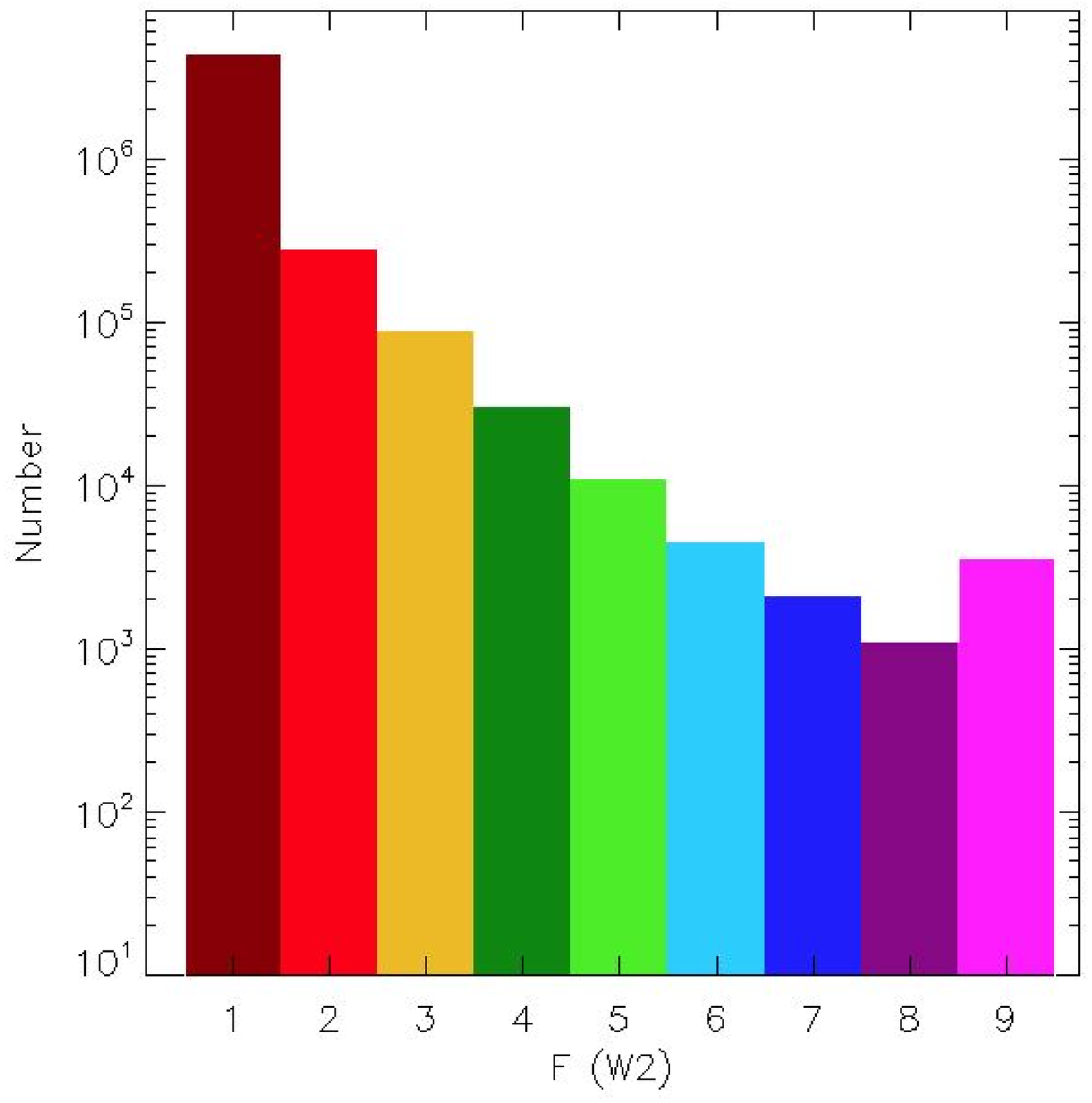}
  \caption[]  
{As with Figure~\ref{fig:w1f}, but for W2.}
  \label{fig:w2f}
\end{figure}

\begin{figure} [hbtp]
 \centering \includegraphics[width=0.7\textwidth]{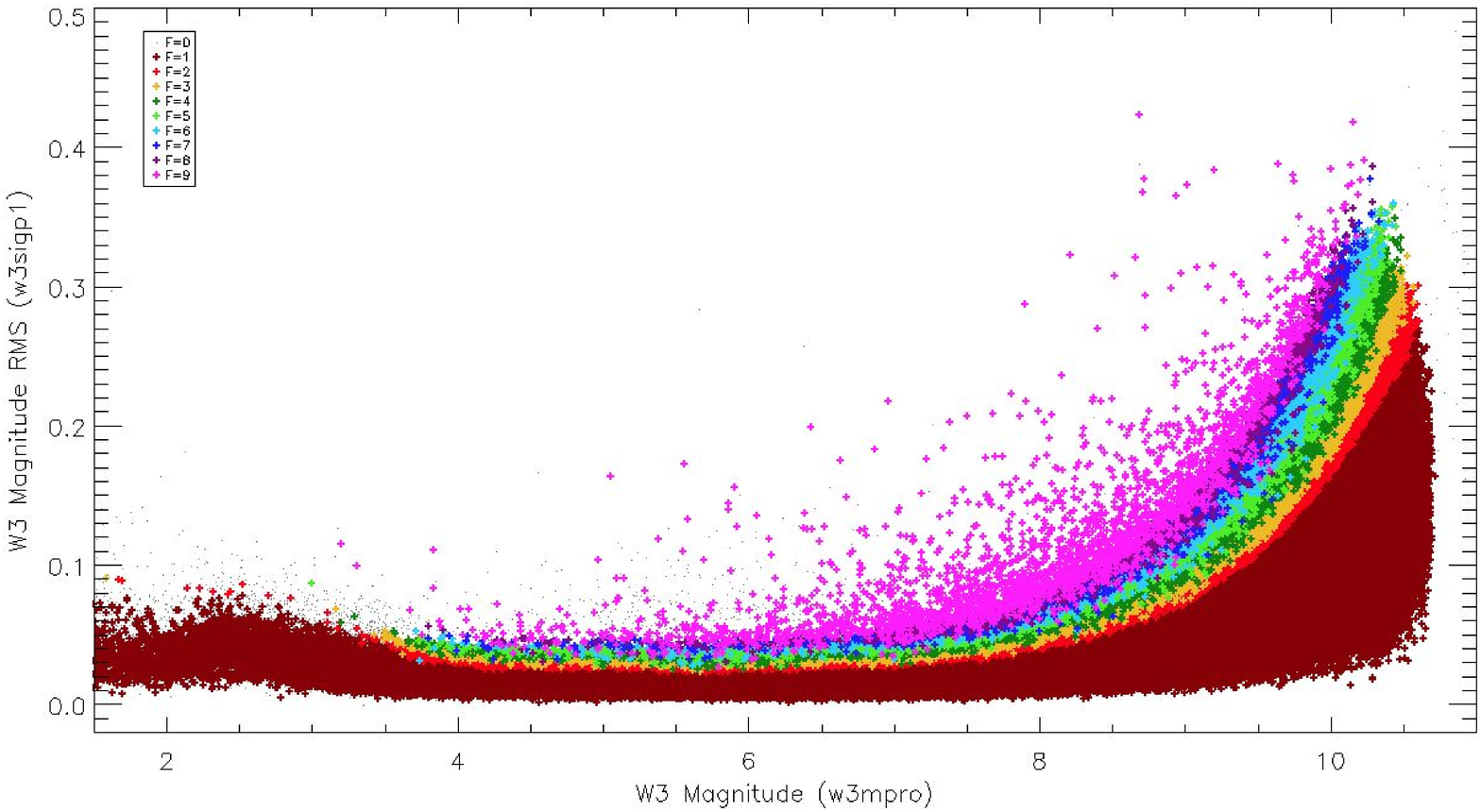}\includegraphics[width=0.3\textwidth]{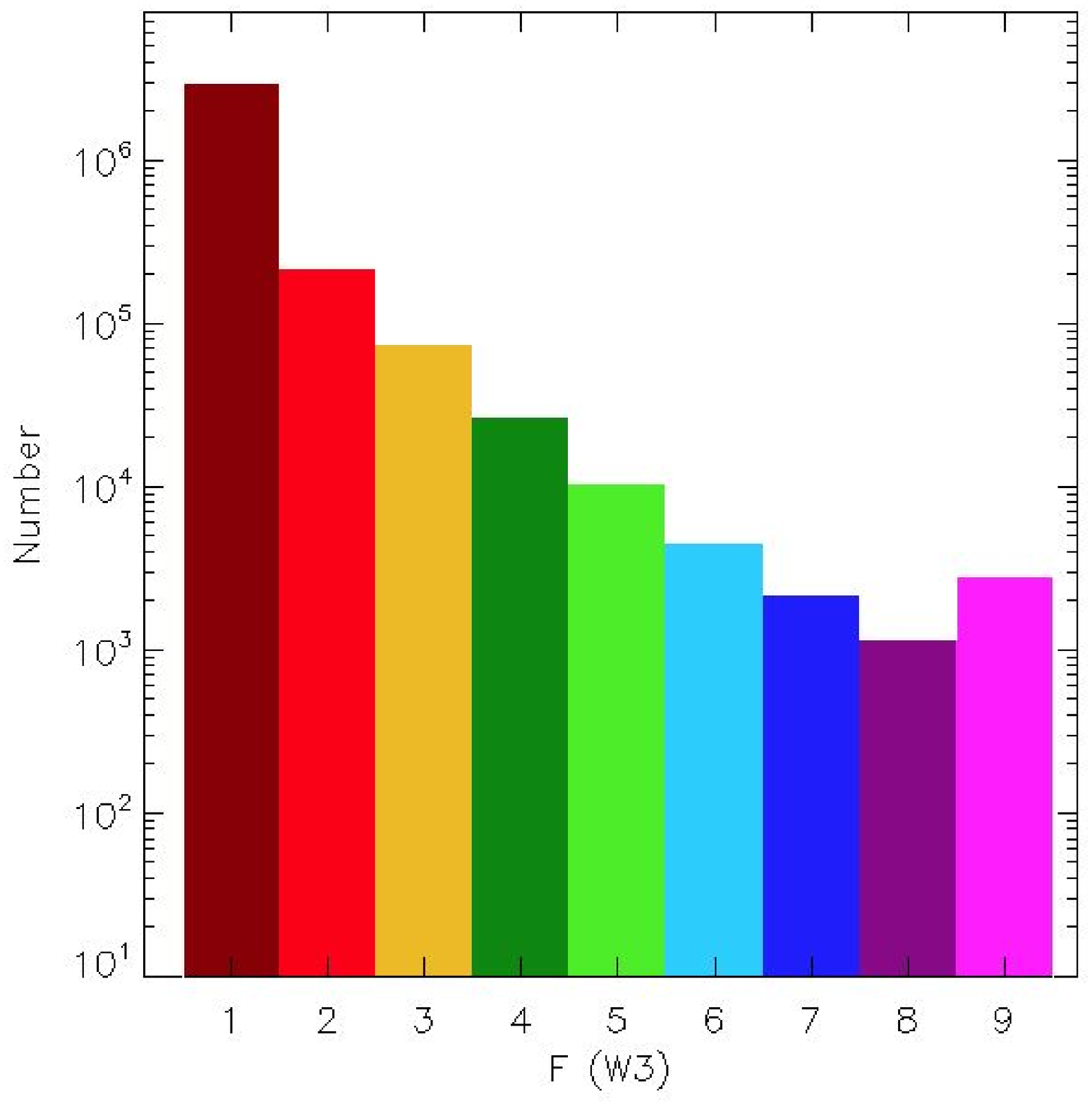}
  \caption[]  
{As with Figure~\ref{fig:w1f}, but for W3.}
  \label{fig:w3f}
\end{figure}

\begin{figure} [hbtp]
 \centering \includegraphics[width=0.7\textwidth]{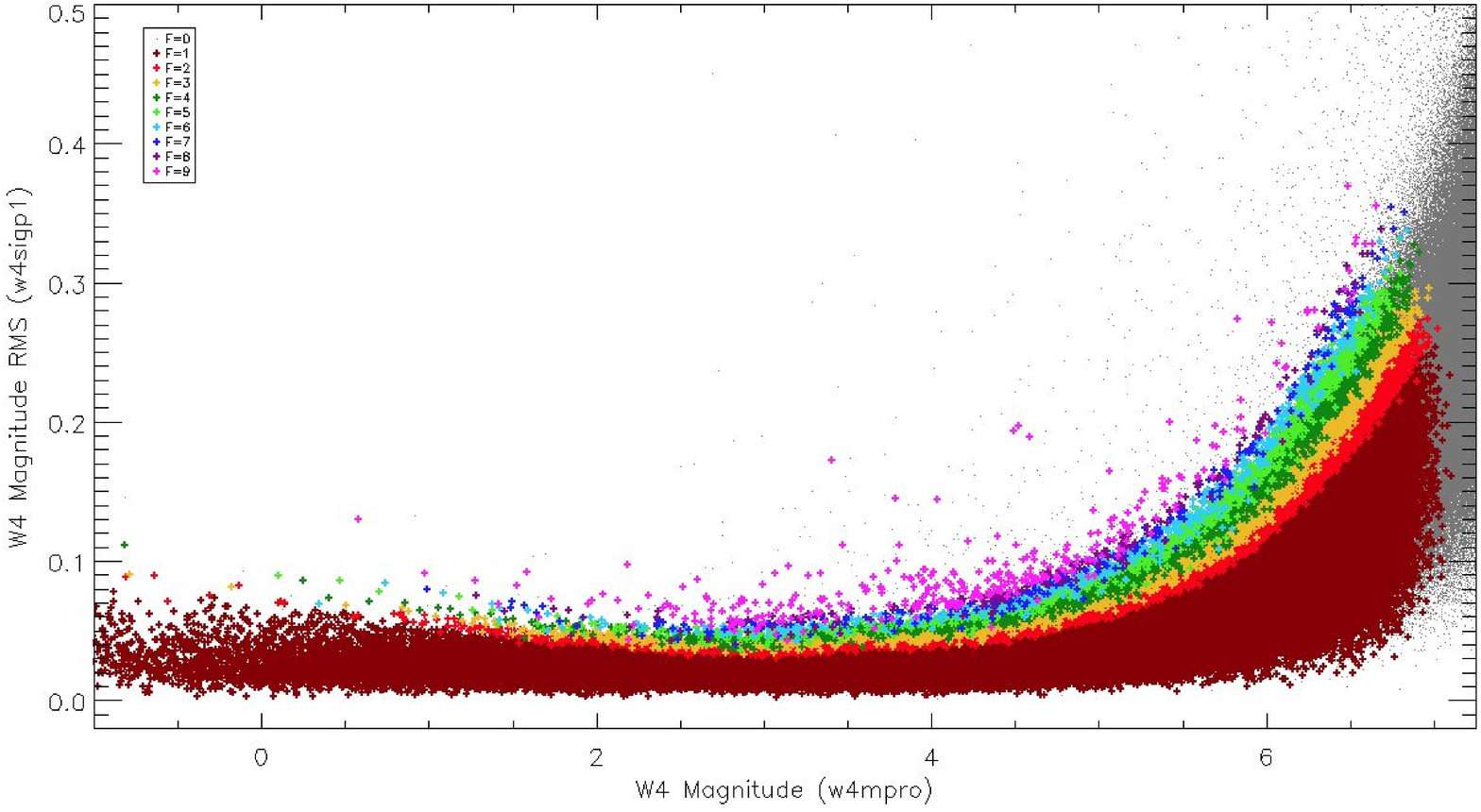}\includegraphics[width=0.3\textwidth]{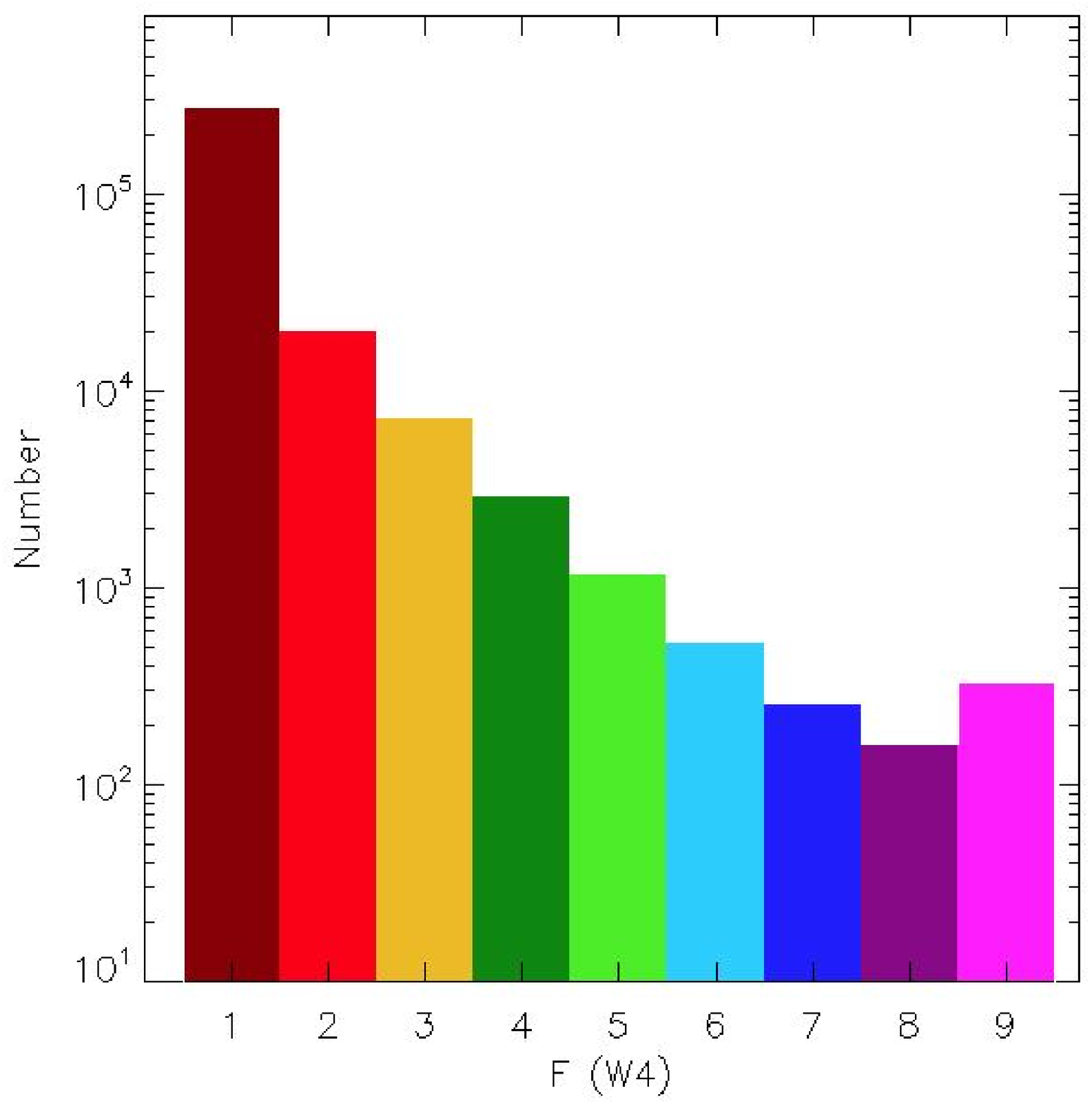}
  \caption[]  
{As with Figure~\ref{fig:w1f}, but for W4.}
  \label{fig:w4f}
\end{figure}

Figures~\ref{fig:w1f}-\ref{fig:w4f} show the results of the method applied to a large subset of the WISE preliminary release data set.  The figures contain $w?sigp1$ as a function of magnitude for each of the four WISE bands.  Each colored cross corresponds to a different value of $F$, with grey dots corresponding to $F=0$.

The values of $F$ cannot be replicated using single-exposure flux measurements.  The method uses the standard deviation of the flux measurements from single-exposures in the multi-frame pipeline.  The fluxes are calculated at the position of the Catalog source, which can differ slightly from the position of the source in the single-exposure images.  Therefore, the fluxes used in the calculation of $w?sigp1$ are not the same as the fluxes in the single-exposure working database.  

There are some known limitations with the flagging algorithm.  Confusion, close companions, and nebulosity can create false variability due to increased noise and positional errors compared to the single-exposure source locations.  Some unflagged artifacts are still present in the preliminary release and contribute to contamination.  The most common of these are sources falling within the halo radius or along a diffraction spike of a parent source that is off-image.  The flagging method is not very sensitive to short-lived, low-amplitude transient events (e.g. flare stars and cataclysmic variables).  A different method is required to detect these types of transients. Despite the largely successful attempt to estimate the single-frame SNR from the coadd SNR, magnitude, and number of frame overlaps, some sources are flagged as highly likely to be variable and have few, if any, single-exposure detections in the given band.  The population distribution of $w?sigp1^2$ does not exactly follow a gaussian or chi-square distribution, but is close to a chi-square where the upper-percentile cut of 65\% for $\sigma$ is a correction.  And finally, artifacts are often flagged too aggressively, leading to many cases where $F=0$ where the light curve is actually quite usable.  Many of these issues will be addressed in the final WISE data release.

\section{Results}
The WISE Preliminary Release Source Catalog consists of over 257 million objects, of which 459,906 are assigned $var\_flg$ values of 7 or greater (Eq. 5) in at least one band.  Only about 5\% of these objects have SIMBAD associations with known variable sources.  The types of periodic variables recovered are dominated by RR Lyr, Algol-type eclipsing binaries, and W UMa binaries.  These types of variables are the most numerous due to the relatively short periods and high amplitudes.  There are few variables seen in WISE bands 3 and 4 mainly due to the lower sensitivity and the declining spectral energy distributions of many stellar sources at these wavelengths, but also because the reddest objects tend to have longer periods than can be observed with one WISE epoch (e.g. Miras and AGN).  

To extract the light curves of suspected variable sources, the position of each catalog entry is queried against the WISE single-exposure source database via the GATOR query service of the NASA/IPAC Infrared Science Archive (IRSA).  Sources in the single-exposure database are not cross-matched with the multiframe sources, thus the association is done by a narrow position cone search.  A search radius of 1.5 arcsec was used to generate the light curves in this paper.  In crowded regions, or for sources with a close neighbor, it is possible for measurements of a non-target source to be included.  In these cases, one must examine the cases where multiple sources are extracted from the same frame and decide which source is the target object.

\begin{figure} [hbtp]
  \begin{center}\plotone{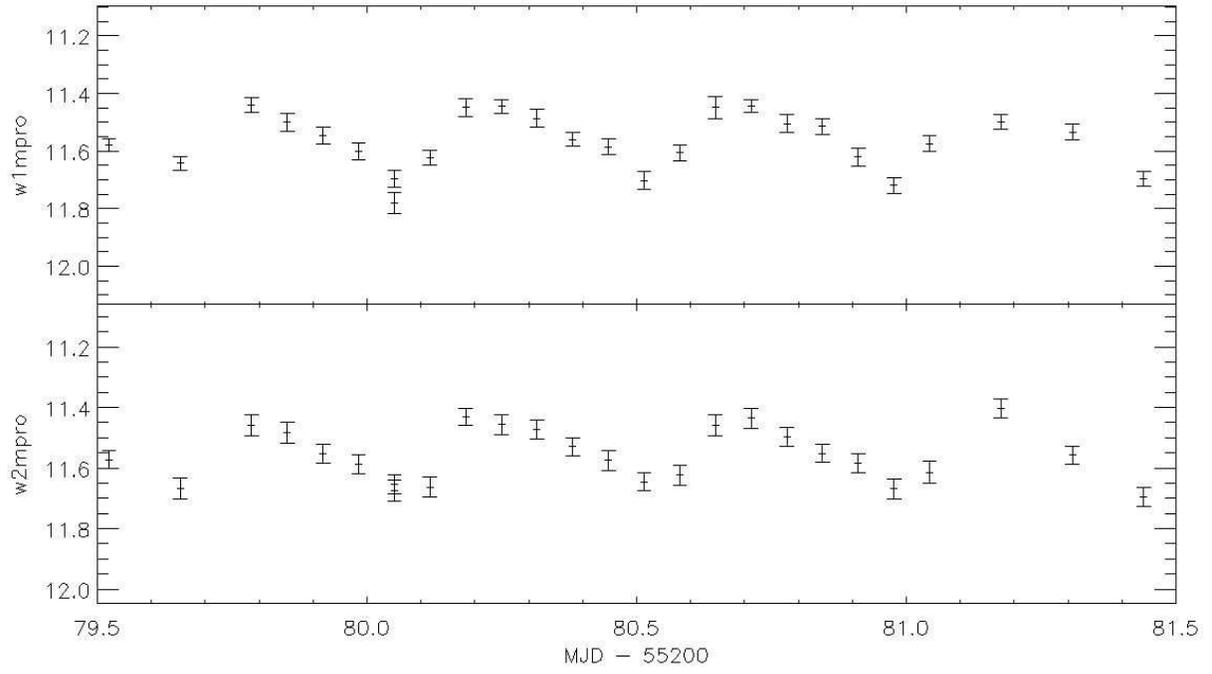}\end{center}
  \caption[]  
{The light curve of the known RR Lyr-type pulsating variable RS Oct.}
  \label{fig:rrlyrlc}
\end{figure}

The typical periods of RR Lyr variables are between 0.2 and 1.0 day, thus more than one complete cycle is usually observed by WISE.  An example light curve of an RR Lyr is given in Figure~\ref{fig:rrlyrlc}.  There are many similar light curves available in the WISE data, but care must be taken to identify these properly as RR Lyr variables.  A common problem is the similar light curve shapes of W UMa and RR Lyr variables of the RRc subclass.  The RRc subclass objects are believed to be first overtone pulsators and have nearly sinusoidal light curves, which can be confused with other types variables.  The RRab subclass (fundamental mode, asymmetric light curve) is more easily identified.  RR Lyr variables are important distance indicators, and WISE can help extend the metallicity-period-luminosity relation into the mid-IR \citep{sandage06}. 

\begin{figure} [hbtp]
  \begin{center}\plotone{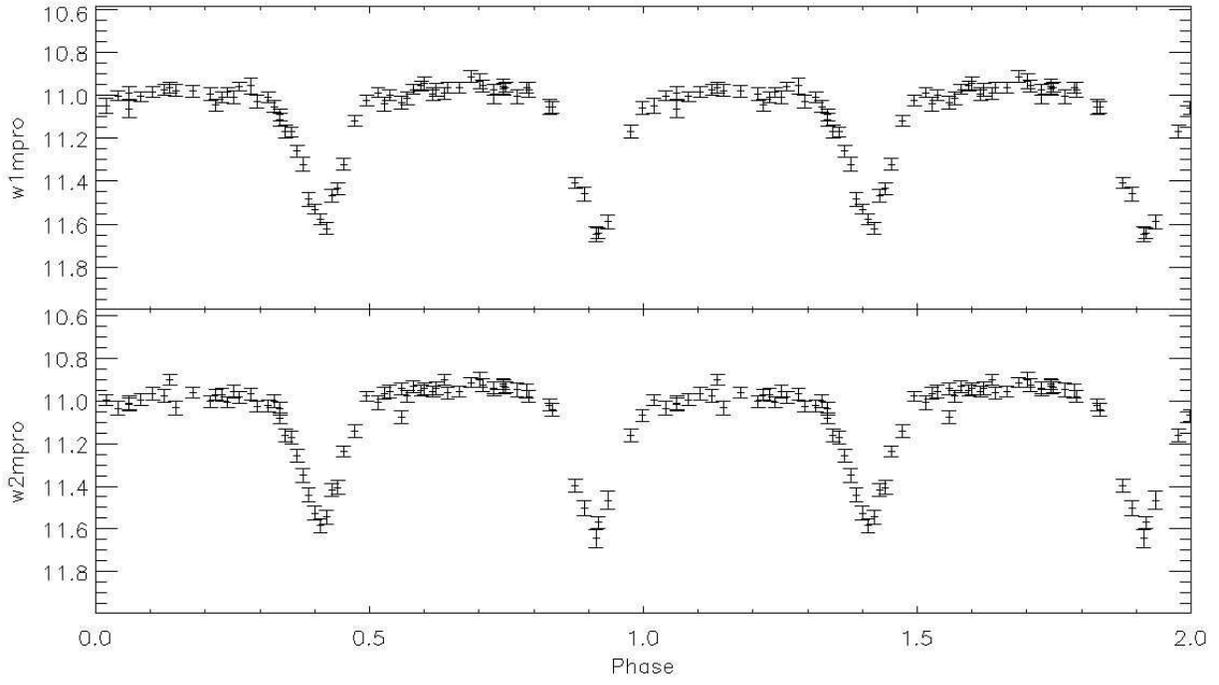}\end{center}
  \caption[Figure 2]  
{The phased light curve of a WISE candidate Algol.  The WISE designation is WISEP J092844.91-732635.8 and is phased to a period of 3.1282 days, determined by chi-square phase dispersion minimization.}
  \label{fig:algol1}
\end{figure}

Algol-type eclipsing binaries typically have longer periods ranging from fractions of a day to many days, thus most WISE Algol light curves are partial, often consisting of just one eclipse.  In these cases, period identification is impossible.  However, for short period Algols and those near the ecliptic poles, where coverage duration is longer, period identification is possible.  Figure~\ref{fig:algol1} is an example of an object at high ecliptic latitude with coverage of a total of 4 eclipses.  The object has not previously been identified as variable and is a candidate Algol-type eclipsing binary with WISE designation WISEP J092844.91-732635.8 and a period of 3.1282 days, determined from chi-square phase dispersion minimization \citep{stellingwerf78}.   

\begin{figure} [hbtp]
  \begin{center}\plotone{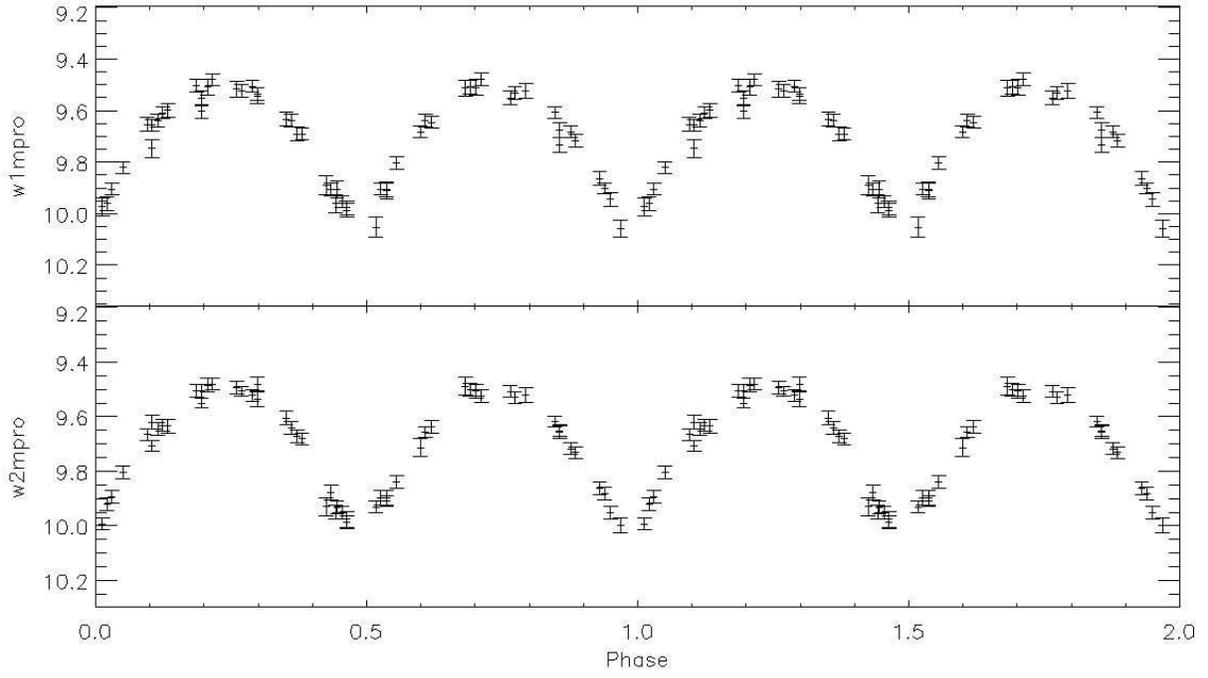}\end{center}
  \caption[Figure 3]  
{Phased WISE light curve of the known W UMa-type contact binary V592 Car.}
  \label{fig:v592car}
\end{figure}

The periods of W UMa binaries are typically between 0.25 and 1.2 days, so WISE often observed several complete cycles.  Many observed W UMa binaries have enough cycles where a phased light curve is possible.  Figure~\ref{fig:v592car} is an example of such a case, where the W UMa binary V592 Car is phased to its orbital period of 0.8011 days.

\begin{figure} [hbtp]
  \begin{center}\plotone{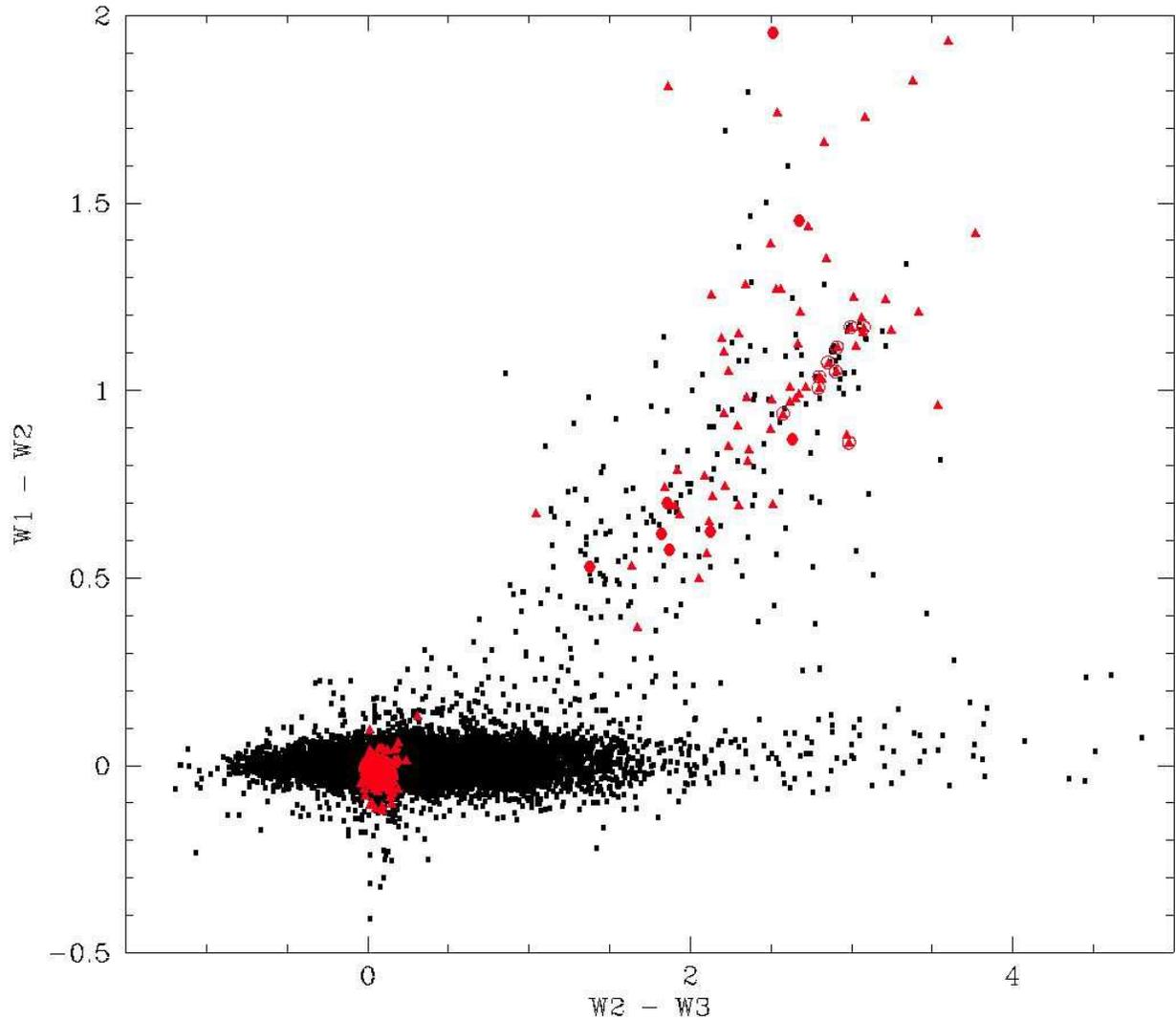}\end{center}
  \caption[Figure 4]  
{Color-color plot of WISE selected variables.  Filled black squares are sources with a variability flag of ``9'' in both bands 1 and 2.  Red symbols are sources with a variability flag of ``9'' in band 1 and 3.  YSO and blazar candidates can be identified by the red population far from the main (stellar) locus of sources.  The black squares in this regime are also likely YSO/blazar candidates, but lack significant variability in W3.  Known blazars are marked with a circled triangle, while known YSOs are marked with a filled red circle.}
  \label{fig:cc}
\end{figure}

\begin{figure} [hbtp]
  \begin{center}\plotone{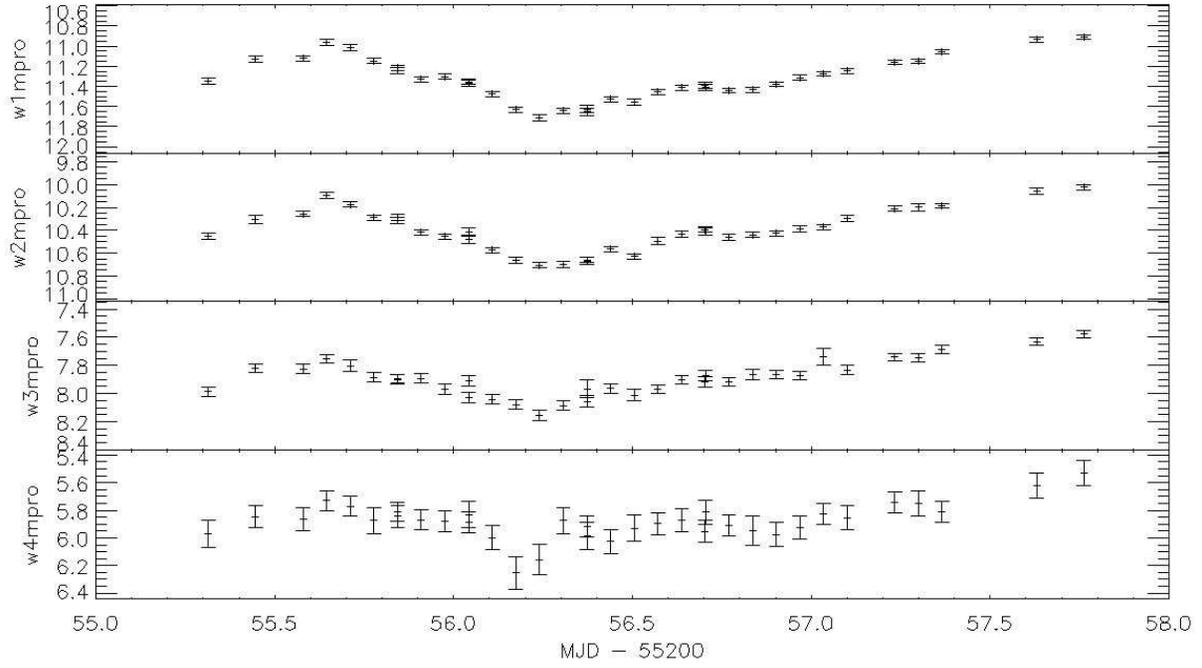}\end{center}
  \caption[Figure 5]  
{WISE light curve of WISEP J053158.61-482736.1 in all four bands.  The non-periodic variable nature of the light curve, the galactic latitude, and the WISE colors are consistent with a blazar.  The object has no previous classification.}
  \label{fig:redcandidate}
\end{figure}

Sources with a high variability likelihood in both W1 and W3 as well
as red colors in both W2$-$W3 and W1$-$W2 are likely blazars or YSOs.
Blazars and YSOs have distinguishing red color and non-periodic
variability in all bands, when available \citep[Figure 12]{Wright10}.
This is clearly seen in the color-color plot of Figure~\ref{fig:cc}.
The dots are sources with a variability flag of ``9'' in both W1 and
W2, whereas the red symbols are sources with a variability flag of
``9'' in W1 and W3.  All of the known blazars and YSOs are in this
color regime, and this simple color cut when combined with the
variability flag can quickly identify blazar and YSO candidates.  The
colors of blazars and YSOs can be quite similar, and distinguishing
between the two can be difficult with WISE data alone.  However,
galactic latitude and observations at other wavelengths can generally
be used to rule out a YSO.  An example of such a candidate is shown in
Figure~\ref{fig:redcandidate}.  The source has the WISE designation
WISEP J053158.61-482736.1 and has WISE colors and light curve
structure consistent with a blazar or variable YSO.  The position of
this source is very close to that of the known flat-spectrum radio
source PMN J0531-4827 \citep{Griffith93}, which suggests strongly that
this object is blazar-like rather than a YSO.  Variability in WISE
bands 3 and 4 is rare, but tends to be dominated by these types of
red, non-periodic variable objects.  Correlation of candidate red WISE
variables with radio source lists may provide an effective means of
discriminating between the two prior to spectroscopic follow-up.

\begin{figure} [hbtp]
  \begin{center}\plotone{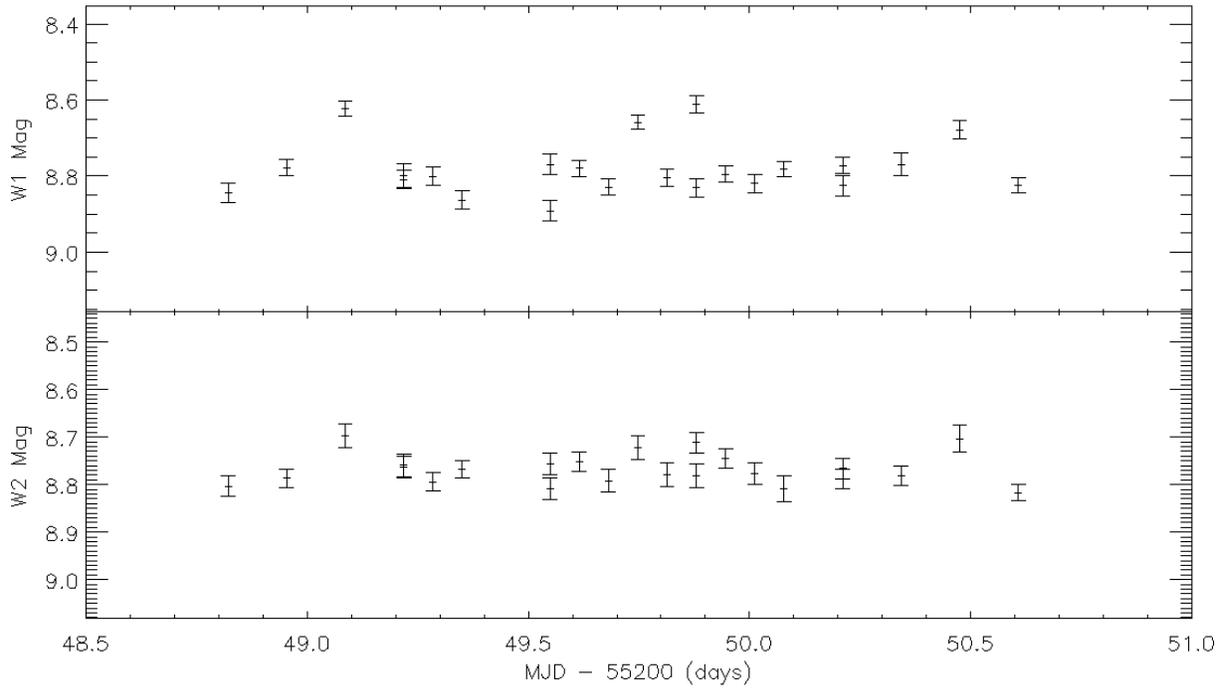}\end{center}
  \caption[]  
{Light curve of a source that is not likely variable.  The source has a variability flag of '7110', and has several close neighbors near its position, which likely affect the PSF photometry.   }
  \label{fig:F7}
\end{figure}

It is difficult to determine the fraction of false variables, as the vast majority of the objects are unknown.  However, approximately 20\% of $F=7$ sources appear to suffer from confusion noise and unflagged artifacts.  Figure~\ref{fig:F7} is an example light curve of a source with $var\_flg$='7110', which is likely a false-positive due to confusion with a nearby source.   The false-positive rate decreases quickly with higher $F$ values, with only about 2\% of $F=9$ sources having the same problem.   Most of the data in the preceding figures are included in the WISE Preliminary Release, with the phased light curve figures containing data not included in the preliminary release, but that will be in the WISE final data release.

\section{Summary} 
We have described the method used to identify nearly 460,000 objects as significantly variable in at least one band in the WISE Preliminary Release Catalog.  The method uses the ensemble statistics available during the multi-frame processing, and assigns a numerical value to each source and band that corresponds to the probability that the source is consistent with flux variations seen in the bulk ``non-variable'' population within the same magnitude bin.  While there are some limitations, the method is practical and effective, given the format and generation of the WISE data. For the WISE All-Sky Release, variability flagging has been further improved, incorporating band-to-band cross correlations, and contains over 6.8 million objects that are significantly variable. 

The objects identified as significantly variable are biased toward periodic variables, with RR Lyr and Algol-type variables being the most common.  Light curves of variables with short periods and/or high depth of coverage can be phased, and period finding or period refinement is possible in these cases.  WISE band 3 variables contain many AGN and YSO sources, which we show can be separated from many other types of variables by a simple color selection.  WISE provides the best all-sky variability data in the mid-IR to date, and will serve the community well.

This publication makes use of data products from the The Wide-field
Infrared Survey Explorer, which is a joint project of the University
of California, Los Angeles, and the Jet Propulsion
Laboratory/California Institute of Technology, funded by the National
Aeronautics and Space Administration.  Long-term archiving and access
to the WISE single-exposure database is funded by NEOWISE, which is a
project of the Jet Propulsion Laboratory/California Institute of
Technology, funded by the Planetary Science Division of the National
Aeronautics and Space Administration.  This research has made use of
the NASA/ IPAC Infrared Science Archive, which is operated by the Jet
Propulsion Laboratory, California Institute of Technology, under
contract with the National Aeronautics and Space Administration.

The authors would like to thank the anonymous referee for valuable
comments leading to an improved evaluation of the approximation errors
affecting the variability indicator.

\appendix
\section{Appendix: Sources of Approximation Error in the Variability Indicator}
Photometric errors are often well described by Gaussian statistics, and this is the case for WISE fluxes measured in DN, i.e., data numbers proportional to flux. For this case, the probability density function describing the distribution of possible flux measurements of a source with true flux $F$ is

\begin{equation}
 p_f(f) = \frac{e^{\frac{-(f-F)^2}{2\sigma_f^2}}}{\sqrt{2\pi}\sigma_f}
\end{equation}

Astronomical magnitude $m$ is related to flux $f$ according to

\begin{equation}
  m = -2.5\log_{10}\left( \frac{f}{f_0} \right) = m_0 - 2.5\log_{10}(f)
\end{equation}
where $m_0$ is a magnitude zero point corresponding to a reference flux $f_0$ that is in the same units as $f$. The second form of the relationship is the one most often encountered in practice, and so we will use that. Note that in taking the logarithm of the flux, we are implicitly clipping the domain to $f > 0$.

Since $m$ is a function of the random variable $f$, we can use standard theory of functions of random variables to obtain the probability density function for $m$,

\begin{equation}
  p_m(m) = \frac{\ln(10)10^{(\frac{m_0-m}{2.5})}e^{-\frac{\left(10^{\frac{m_0-m}{2.5}}-F \right) ^2}{2\sigma^2_f}}}{2.5\sqrt{2\pi}\sigma_f}
\end{equation}

For relatively high SNR, the shape of this density function is practically indistinguishable from a Gaussian, but for low SNR, it differs very significantly, being positively skewed and leptokurtic. Since most sources detected in a survey are in the neighborhood of the detection threshold, they dominate the statistics, and their large deviations from Gaussian behavior cause the variability indicator based on a chi-square formula to have large tails that distort the $Q$ distribution relative to a true chi-square.

The variability indicator is based on a reconstruction of the expression

\begin{equation}
 \Psi \equiv \sum^N_{i=1}\frac{(m_i-\bar{m})^2}{\sigma^2_i}
\end{equation}

This expression is chi-square if and only if the $m_i$ are Gaussian and uncorrelated. We denote this function $\Psi$ instead of $\chi^2$ because we have seen above that the $m_i$ are not Gaussian for low SNR, and now we examine the error incurred by ignoring their correlations, which they develop to some extent because of the flux estimation algorithm. Here we consider only the latter effect and take the actual distribution to be Gaussian. That allows us to use the general multivariate density function for $N$ correlated Gaussian random variables, the vector $\boldsymbol{m_N}$

\begin{equation}
p_N(\boldsymbol{m_N}) = \frac{e^{-\chi^2_N / 2}}{(2\pi)^{N/2}\sqrt{|\Omega_N|}}
\end{equation}
where $\Omega_N$ is the covariance matrix, and $|\Omega_N|$ is its determinant. Using $\rho_{ij}$ to denote the correlation
coefficient for $m_i$ and $m_j$, the covariance matrix and $\chi^2_N$ are

\begin{equation}
\Omega_N = \begin{pmatrix} \sigma^2_1 & \rho_{12}\sigma_1\sigma_2 & \cdots & \rho_{1N}\sigma_1\sigma_N \\
  \rho_{21}\sigma_2\sigma_1 & \sigma^2_2 & \cdots & \rho_{2N}\sigma_2\sigma_N \\
  \vdots & \vdots & \ddots & \vdots \\
  \rho_{N1}\sigma_N\sigma_1 & \rho_{N2}\sigma_N\sigma_2 & \cdots & \sigma^2_N \end{pmatrix}
\end{equation} 

\begin{equation}
\chi^2_N = \sum^N_{i=1}\sum^N_{j=1}w_{ij}(m_i-\bar{m})(m_j-\bar{m})
\end{equation}
where the $w_{ij}$ are the elements of the inverse of the covariance matrix.

For any value of $N$, we can compute the mean and variance of $\Psi$ by formally performing the corresponding moment integrals using the joint density function above. The result is that the expectation value of $\Psi$ is $N$, the same as for a chi-square random variable, independently of any correlations, which is obvious from the fact that the definition of $\Psi$ consists of a sum of $N$ terms, each of whose expectation is unity, and the variance of $\Psi$ is

\begin{equation}
\rm{var}(\Psi) = 2N + 4\sum^{N-1}_{i=1}\sum^N_{j=i+1}\rho^2_{ij}
\end{equation}

If the correlations are all zero, then this variance reduces to $2N$, and $\Psi$ is chi-square. Otherwise, any correlation can only increase the variance, causing $Q(\Psi)$ to approach zero more slowly than the $Q$ distribution for chi-square with correlations properly taken into account (e.g. Equation A7).

\clearpage

\clearpage

\clearpage

\clearpage

\clearpage

\clearpage

\clearpage

\clearpage

\clearpage

\clearpage

\clearpage

\begin{deluxetable}{lp{12cm}}
\tablecaption{Selected WISE Catalog Column Definitions\label{tab:defs}}
\tablecolumns{2}

\tablehead{
\colhead{Column Name} &
\colhead{Definition}
}

\startdata
w?mpro & The profile-fit photometetric measurement in magnitudes for the specified band. \\
\\
w?sigmpro &  The profile-fit photometric measurement uncertainty in mag units for the specified band. \\
\\
w?snr & The profile-fit measurement signal-to-noise ratio. \\
\\
w?rchi2 & Reduced $\chi ^2$ of the specified band's profile-fit photometry measurement. \\ 
\\
w?sigp1 & Standard deviation of the population of the specified band's fluxes measured on the individual frames covering the source, in magnitudes. \\
\\
w?m & The number of usable frames containing the source's coordinates in the specified band. \\
\\
w?nm & The number of detections with $SNR > 3$ in the specified band. \\
\\
cc\_flags & A four character string, one character per band [W1/W2/W3/W4], that indicates that the photometry and/or position measurements of a source may be contaminated or biased due to proximity to an image artifact.  Non-zero characters denote possible artifacts. \\
\\
na & Active deblending flag.  ``0'' indicates the source is not actively deblended while ``1'' indicated the source is actively deblended. \\
\\
nb & Number of PSF components used simultaneously in the profile-fitting for this source. \\
\\
var\_flg &  The variability flag is a four-character string, one character per band, in which the value for each band is related to the probability that the source flux measured on the individual WISE exposures was not constant with time.  A value of zero indicates insufficient or inadequate data to determine a variability probability, while values of 1-9 indicate increasing likelihood of variability. \\
\enddata
\end{deluxetable}

\end{document}